\documentclass[referee,a4paper]{raa}            

\usepackage{graphicx,times}             
\usepackage{multirow}
\usepackage{txfonts}
\usepackage{color}
\usepackage{threeparttable}
\newcommand{\tabincell}[2]{\begin{tabular}{@{}#1@{}}#2\end{tabular}}

\usepackage{natbib}
\usepackage{epsfig}
\usepackage{float}
\usepackage{bm}

\begin{document}

   \title{Numerical Simulations of Solar Energetic Particle Event Timescales 
Associated with ICMES
}

   \volnopage{Vol.0 (200x) No.0, 000--000}      
   \setcounter{page}{1}          

   \author{S.-Y. Qi
      \inst{1,2}
   \and G. Qin
      \inst{3,1,2}
   \and Y. Wang
      \inst{3}
   }

   \institute{State Key Laboratory of Space Weather,  National Space Science 
Center, Chinese Academy of Sciences, Beijing 100190, China\\
        \and
             College of Earth Sciences, University of Chinese Academy of 
Sciences, Beijing 100049, China\\
        \and
             School of Science, Harbin Institute of Technology, Shenzhen, 518055, China; 
             {\it qingang@hitsz.edu.cn}\\
   }

   \date{Received~~2009 month day; accepted~~2009~~month day}

\abstract{ Recently, S.W. Kahler studied the solar energetic particle (SEP) 
event timescales associated with coronal mass ejections (CMEs) 
from spacecraft data analysis. They obtained different timescales of SEP 
events, such as $TO$, the onset time from CME launch to SEP onset, $TR$, the 
rise time from onset to half the peak intensity ($0.5I_{p}$), and $TD$, the 
duration of the SEP intensity above $0.5I_{p}$. In this work, we solve SEPs 
transport equation considering ICME shocks as energetic particle sources. 
With our modeling assumptions, our simulations show similar results to 
Kahler's spacecraft data analysis that the weighted average of $TD$ increases 
with both CME speed and width. Besides, from our simulation results, we suggest $TD$ is directly dependent on CME speed, but not dependent on CME width, which were not achieved from the observation data analysis.
}

   \authorrunning{S.-Y. Qi, G. Qin \& Y. Wang }            
   \titlerunning{Numerical Simulations of Solar Energetic Particle Event Timescales 
Associated with ICMES }  

   \maketitle
\section{Introduction}           
\label{sec:intro}

Solar energetic particle (SEP) events could be mainly divided into two classes through duration and intensity. The short-duration and low-intensity events, which are called impulsive events, are considered to be produced by solar flares. On the other hand, the longer duration and higher intensity ones, which are called gradual events, are considered produced by coronal and interplanetary shocks driven by coronal mass ejections (CMEs). It is interesting to study the relationship between gradual SEP event properties and the characteristics of the associated CMEs. With the first-order Fermi acceleration mechanism \citep{Zank2000JGR...10525079Z} introduced an onion shell model using a one-dimensional hydrodynamic code for the evolution of the CME-driven shock in the Parker interplanetary magnetic field (IMF). The model is valid only in strong shocks due to Bohm diffusion coefficient used, so \cite{Rice2003JGRA..108.1369R} modified it to be usable in arbitrary strengths. In addition, \cite{Li2003JGR...1081082L} studied the transport of SEPs with their onion shell acceleration model considering particles pitch angle scattering without perpendicular diffusion. In their model, charged particles' pitch angle diffusion is not considered between two consecutive pitch angle scatterings. Furthermore, \cite{Verkhoglyadova2009APJ..693.894V, Verkhoglyadova2010JGR..115.A12103V} adopted this model to study individual SEP events caused by CME shocks, their simulation results can fit well with spacecraft observations for different elements. On the other hand, considering that the interplanetary coronal mass ejection (ICME) shocks can continuously accelerate SEPs when propagating outward, \citet{Kallenrode1997jgr...102.22311K, Kallenrode2001JGR10624989K} treated the ICME shock as a moving particle source. And the model was adopted in a numerical code \footnote{Hereafter, we denote the code as Shock Particle Transport Code, SPTC.} by \citet{Wang2012ApJ...752..37W} to study ICME driven shock accelerated 
particles' transport in three dimensional solar wind and IMF including both parallel and perpendicular diffusion coefficients. Furthermore, under varying perpendicular diffusion and shock acceleration strength, \citet{QinEA13} reproduced the reservoir phenomenon with SPTC numerical simulations. In addition, with the same numerical modeling, \citet{WangAQin15} researched the gradual SEP events spectra forcusing on the spatial and temporal invariance. Finally, \citet{QinAWang15} compared the simulation results from SPTC with the multi-spacecraft ($Helios$ $1$, $Helios$ $2$, and $IMP$ $8$) observations during a 
gradual SEP event, and they obtained the SPTC simulations which best fit the SEP event observed by spacecraft located in different space.

To investigate the relationship between SEP event properties with the associated CMEs, \citet{Ding2014ApJ...793L..35D} studied the interaction of two CMEs erupted nearby during a large SEP event by multiple spacecraft observations with the graduated cylindrical shell model. And they obtained the solar particle release time and path length which indicated the necessary influence of the "twin-CME" \citep{Li2012SSRv..171..141L, Temmer2012ApJ...749...57T} on the SEP event. 

Because of the huge damage caused by SEPs, the study of peak intensities of SEPs 
becomes very important. \citet{Ding2015ApJ...812..171D} presented the new
observation results of peak intensity with Fe/O ratio, which indicate the role of seed population in extremely large
SEPs. \citet{Reinard2006asr38...38.480R}
studied the dependence of the occurrence and peak intensities of SEP events with CME 
properties thoroughly using databases of the LASCO/$SOHO$ CMEs and the $GOES$ $E>10$
MeV protons. Besides peak intensities, timescales are another very important 
property of SEPs which could make contribution to both space weather forecasting and
understanding of the SEP injection profiles and propagation characteristics. 

In order to study the properties and associations of SEP events, 
\cite{Cane2010JGR...115.A08101C} compared SEPs with flares and CMEs of $280$ solar 
proton events which extended above $25$ MeV occurred from $1997$ to $2006$ by 
near-Earth spacecraft. They divided the events into $5$ groups according to the 
ratios $e/p$ and Fe/O at event onset. Their results suggested that SEP event 
occurrence and peak intensities are more likely to be associated with faster and 
wider CMEs, especially with western CME source regions. Furthermore, 
\cite{Pan2011SoPh...270.593P} investigated SEP timescales, such as the SEP onset 
time, the SEP rise time, and the SEP duration. With an ice-cream cone model, 
\cite{Pan2011SoPh...270.593P} studied LASCO/$SOHO$ observation data of $95$ CMEs 
associated with SEP events during $1998-2002$, and came to conclusions that the SEP 
onset time has no significant correlation with the CME speed, nor with the CME 
width. They also suggested that the SEP rise time and the SEP duration have significantly positive 
correlations with the radial speed and angular width of the associated CMEs unless 
the events are not magnetically well connected to the Earth.

\cite{Kahler2013ApJ...769.110K} did a research on the relationship between the 
EPACT/$Wind$ $20$ MeV SEP events timescales and their associated CME speed and 
widths observed by LASCO/$SOHO$. In \cite{Kahler2013ApJ...769.110K}, $217$ SEP 
events observed in a solar cycle during the period 1996-2008 were used. They 
defined the three characteristic times of the SEP events. The time from inferred 
CME launch at $1$ $R_{\odot}$ to the time of the $20$ MeV SEP onset at $Wind$ 
was denoted as $TO$. The time from SEP onset to the time the intensity reached half of 
the peak value ($0.5I_{p}$)was denoted as $TR$. And the time during which the intensity
was above $0.5I_{p}$ was denoted as $TD$. From their results, they found that CME 
speed and width were of significant correlation and it is not easy to interpret the 
contribution of CME speed and width to timescales separately. Therefore, they 
suggested that faster and wider CMEs which drive shocks and accelerate SEPs over 
longer times would thus produce the longer SEP timescales $TR$ and $TD$.

In this paper, with the data used in the analysis of 
\citet{Kahler2013ApJ...769.110K}, we study the CME timescales by numerical 
simulations with the SPTC, and we compare our results with that of 
\citet{Kahler2013ApJ...769.110K}. In section 2, we present the model. In section 3, 
we present the data analysis. In section 4, we show our results. In section 5, we 
present the conclusions and discussion.

\section{MODEL}
\label{sctn:model}

We model the transport of SEPs by following previous research
\citep[e.g.,][]{Qin2006JGRA..11108101Q,Zhang2009ApJ...692..109Z}.
The three-dimensional focused transport equation is written as
\citep{Skilling1971ApJ...170..265S,Schlickeiser2002cra..book.....S,
Qin2006JGRA..11108101Q,Zhang2009ApJ...692..109Z}
\begin{eqnarray}
  \frac{{\partial f}}{{\partial t}} = 
  \nabla\cdot\left(\boldmath{\kappa_\bot}\cdot\nabla f\right)
+ \frac{\partial }{{\partial \mu }}\left(D_{\mu \mu }\frac{{\partial f}}{{\partial \mu }}\right) \nonumber 
- \left(v\mu \bm{\mathop b\limits^ \wedge}+ \bm{V}^{sw}\right)\cdot \nabla f \\
+ p\left[ {\frac{{1 - \mu ^2 }}{2}\left( {\nabla  \cdot \bm{V}^  {sw}  
-\bm{\mathop b\limits^ \wedge  \mathop b\limits^ \wedge } :\nabla\bm{V}^{sw} } \right) 
+\mu ^2 \bm{\mathop b\limits^ \wedge  \mathop b\limits^ \wedge}  :\nabla \bm{V}^{sw} }\right]\frac{{\partial f}}{{\partial p}} \nonumber \\
- \frac{{1 - \mu ^2 }}{2}\left[ { - \frac{v}{L} + \mu \left ( {\nabla  \cdot \bm{V}^{sw} 
- 3\bm{\mathop b\limits^ \wedge  \mathop b\limits^\wedge}  :\nabla \bm{V}^{sw} }
\right)} \right]\frac{{\partial f}}{{\partial \mu }},\label{dfdt}
\end{eqnarray}
where $f(\bm{x},\mu,p,t)$ is the gyrophase-averaged distribution function, $\bm{x}$ 
is the position in a non-rotating heliographic coordinate system, $\mu$ is the 
particle pitch-angle cosine, $p$ is the particle momentum, $v$ is the particle 
speed, $t$ is the time, ${{\bm{\kappa }}_ \bot }$ and $D_{\mu \mu }$ are the 
particle perpendicular and pitch-angle diffusion coefficients, respectively,
$\bm{V}^{sw}=V^{sw}\bm{\mathop r\limits^ \wedge}$ is the solar wind velocity which 
is in the radial direction, and $L=\left(\bm{\mathop b\limits^ \wedge}
\cdot\nabla\ln B_0\right)^{-1}$ is the magnetic focusing length determined by the 
magnitude of the background magnetic field $B_0$ and the unit vector along the local 
magnetic field $\bm{\mathop b\limits^ \wedge}$. 
In the equation (\ref{dfdt}), almost all important transport effects are included,
i.e., perpendicular diffusion (1st term in RHS), pitch angle diffusion (2nd term
in RHS), particle streaming along field line and solar wind flowing in the IMF 
(third term in RHS), adiabatic cooling in the expanding solar wind (4th term in 
RHS), and magnetic focusing in the diverging IMF (5th term in RHS). Here, the drift 
effects are neglected for lower-energy SEP transport in the inner heliosphere. Also 
the IMF is modeled with the Parker field.

By following \cite{BurgerEA2008apj..674.511}, diffusion coefficients are determined.
We set the perpendicular diffusion coefficient from the nonlinear guiding center 
(NLGC) theory \citep{Matthaeus2003ApJ..590.53M} approximated with the analytical 
form according to \citet{Shalchi2004ApJ...616..617S,Shalchi2010Ap&SS.325...99S},
\begin{equation}
{{\bm{\kappa }}_ \bot } = v l_d^{2/3}\lambda _\parallel^{1/3}
\left({ {\bf{I}} - \mathop {\bf{b}}\limits^ \wedge \mathop {\bf{b}}
\limits^ \wedge  } \right),\label{kappa_per}
\end{equation}

where $l_d$ is a parameter to control the value of the perpendicular diffusion 
coefficient. For simplicity, ${{\bm{\kappa }}_ \bot }$ is set to be independent of 
$\mu$ with the assumption that particle pitch-angle diffusion is much faster than 
perpendicular diffusion, but generally $\mu$ dependent perpendicular diffusion 
coefficient should be used \citep[e.g.,][]{QinAShalchi14a}.

The parallel particle mean free path (mfp) $\lambda _\parallel$ is written as 
\citep{Jokipii1966ApJ...146..480J,HasselmannAWibberenz68,Earl74}
\begin{equation}
\lambda _\parallel = \frac{{3\upsilon}}{8}\int_{ - 1}^{ + 1}{\frac{{(1 - 
\mu ^2 )^2 }}{{D_{\mu \mu } }}d\mu},\label{lambda_parallel_1}
\end{equation}
and parallel diffusion coefficient $\bm{\kappa}_\parallel$ can be written as 
$\bm{\kappa}_\parallel=v\lambda_\parallel/3$.

We follow \citet{Beeck1986ApJ...311..437B} and  \citet{Teufel2003AA..397.15T} to 
model the pitch angle diffusion coefficient 
\begin{equation}
 D_{\mu \mu }(\mu)= Gv R_{L}^{s-2} \left\{ {\left. {\left| \mu \right.}
 \right|^{s-1}+ h} \right \} \left( {1 - \mu ^2 } \right),\label{D_mu_mu}
\end{equation}
where $G$ is a parameter to control the value of $D_{\mu \mu }(\mu)$, 
$v$ is the particle speed, $R_{L}=pc/(\left. {\left| q\right.}\right|B_0)$ 
is the particle Larmor radius. Here, a larger value of $h=0.01$ is chosen 
for non-linear effect of pitch angle diffusion at $\mu=0$ in the solar
wind \citep{QinAShalchi09, qin2014detailed}.

To model the particle injection, the shock is treated as a moving SEP source with 
the boundary condition \citep{Kallenrode1997jgr...102.22311K}:
\begin{equation}
f_b = a \delta (r-v_{s}t) {\left({ \frac{r}{r_{c}}}\right)}
^{\alpha} \exp \left[{- \frac{ \left| {\phi (\theta ,
\varphi )} \right|}{\phi_{c}(p)}}\right]p^{-\gamma}H(\phi_{s}-\left|{\phi(\theta,
\varphi )} \right|) 
\end{equation}
where $\alpha$ and $\phi_c$ are the shock acceleration strength parameters. We 
assume $\phi_c$ as a constant, but $\alpha$ as a function of shock speed, e.g.,
we set
\begin{equation}
\alpha=\left\{\begin{array}{l@{\quad}l}
-3.5&  \rm{if} \,\,\emph v_s <\emph v_1\\
v_s/v_0-5&  \rm{if} \,\, \emph v_1\le \emph v_s\le \emph v_2\\
-2& \rm{if} \,\, \emph v_s > \emph v_2\\
\end{array}\right.
\end{equation}
where $v_0=500$ km s${}^{-1}$, $v_1=750$ km s${}^{-1}$, and $v_2=1500$ km s${}^{-1}$. $\phi(\theta,\varphi)$ 
is the angle between source center and any point of particle injection 
$(\theta,\varphi)$. $\gamma$ is the spectral index of source particles. In the 
simulations, we inject energetic particle shells with small space intervals 
$\Delta r$. $H(x)$ is the Heaviside step function, with $\phi_{s}$ being the half
angular width of the shock. A more detailed description of the shock model of our 
simulations can be referred to \cite{Wang2012ApJ...752..37W}.

The transport equation (\ref{dfdt}) is solved by a time-backward Markov stochastic 
process method \citep{Zhang1999ApJ...513..409Z} in the simulations. And the detailed 
description of the method can be referred to \citet{Qin2006JGRA..11108101Q}. As 
mentioned in section \ref{sec:intro}, our numerical code of transport of energetic 
particles with the CME driven shock as a moving particle source is denoted as Shock 
Particle Transport Code, i.e., SPTC.

\section{DATA  ANALYSIS}
\label{sect:data}

We investigate $20$ MeV proton intensity-time profiles of SEP events during 1996 
to 2008 with their associated CMEs. In particular, the SEP data is from  EPACT 
(the Energetic Particles: Acceleration, Composition, and Transport)
\citep{1995SSRv...71..155V} experiment on the $Wind$ spacecraft, and the 
information of their related CMEs is observed by $SOHO$ (the $Solar$ $and$ 
$Heliospheric$ $Observatory$ mission) LASCO (Large Angle and Spectrometric 
Coronagraph) \citep{1995SoPh..162..357B}. Of the total $217$ SEPs during this 
period \citep{Kahler2013ApJ...769.110K}, we study $204$ SEPs whose CME parameters 
are available. In addition, for each event, the CME solar source is determined by 
flare location, and the speed ($v_{CME}$) and width ($W_{CME}$) of CME are obtained from \citet{Kahler2013ApJ...769.110K}. 

\subsection{Parameter Selection}

For the grouping and selecting data, we follow the method suggested by 
\citet{Kahler2013ApJ...769.110K} dividing the $204$ events into five longitude 
ranges with about $41$ events each, and subdividing each longitude range into 
several groups sorted on $v_{CME}$ and $W_{CME}$, respectively. The median values 
of longitude, $v_{CME}$, and $W_{shock}$ in each group are used as the 
characteristic values.

From data analysis of spacecraft observations, it is not easy to identify SEP onset 
time accurately which is usually covered by the background of intensity. Therefore, 
in this work, we only focus on the variation of TD with $v_{CME}$ and $W_{CME}$. To compare with the observation,
we obtain the data analysis results of variation of TD with $v_{CME}$ and $W_{CME}$ 
from \citet{Kahler2013ApJ...769.110K} as shown in Table \ref{tbl:kahler}. From 
Table \ref{tbl:kahler} we can see, we study SEP events with source location 
longitude in three ranges, W$33-$W$60$, W$62-$W$90$, and W$100-$bWL, with median 
values W$48$, W$77$, and W$112$, respectively. Note that bWL indicates sources 
behind the west limbs. In each range of longitude, $TD$ is shown as varying with 
the median values of $v_{CME}$ and $W_{CME}$ by subdividing the range into several 
groups sorted on $v_{CME}$ and $W_{CME}$, respectively.

\begin{table}[!htb]
\begin{centering}
\caption {The data analysis results of variation of TD with $v_{CME}$ and $W_{CME}$ 
from \citet{Kahler2013ApJ...769.110K}.\label{tbl:kahler}}
\begin{tabular} {c|lr|lr|lr} 
\hline
Source Location&\multicolumn{2}{c|}{W$33$-W$60$}&
\multicolumn{2}{c|}{W$62$-W$90$}&\multicolumn{2}{c}{W$100$-bWL}\\
Longitude&\multicolumn{2}{c|} ~ & ~ & ~
\\
\hline
\multirow{5}{*}{\tabincell{c}{TD varying\\ with $v_{CME}$}}&$v_{CME}$ (km/s) &TD (h)
&$v_{CME}$ (km/s)&TD (h)&$v_{CME}$ (km/s)&TD (h)\\
&$450$ & $6.3$ &$650$ & $6.5$ & $620$ & $13.2$ \\
&$800$ & $12.0$ & $1150$ & $9.8$ & $900$ & $14.0$ \\
& $1175$ & $8.8$ & $1450$ & $21.3$ & $1325$ & $12.5$ \\
& $1600$ & $14.5$ & $2100$ & $18.1$ & $1750$ & $17.0$ \\
\hline
\multirow{4}{*}{\tabincell{c}{TD Varying\\ with $W_{CME}$}}&$W_{CME}$ (${}^\circ$)&
TD (h)&$W_{CME}$ (${}^\circ$)&TD (h)&$W_{CME}$ (${}^\circ$)&TD (h)\\
& $77$  & $8.3$&$133$ & $7.8$ &$100$  & $7.5$\\
& $208$ & $11.3$ & $171$  & $15.0$ & $178$  & $13.3$ \\
& $360$  & $15.8$ & $360$  & $16.4$& $360$&$17.2$\\
\hline
\end{tabular}
\end{centering}
\end{table}

In order to study SEP timescales associated with CMEs, we use the SPTC described 
in Section \ref{sctn:model} to simulate the transport of SEPs assuming the CME 
shock as a moving particle source and that the shock nose is in the flare 
direction relative to the solar center. In SPTC, the speed of shock, $v_s$, and 
the width of shock, $W_s$, are needed. While, in the spacecraft data analysis of 
\citet{Kahler2013ApJ...769.110K} the speed and width of CME are used instead. 
To compare the simulation results with the spacecraft data analysis, we need 
a model for relationship between $v_s$ and $v_{CME}$, and that for relationship 
between $W_s$ and $W_{CME}$. Firstly, we assume the speed of CME is the same as 
that of shock, $v_s=v_{CME}$. Secondly, since the width of shock ($W_s$) is 
larger than that of CME ($W_{CME}$),we set, 
\begin{equation}
W_s=\left\{\begin{array}{l@{\quad}l}
W_{CME}+\Delta W &  \rm{if} \,\, \emph W_{CME}< 360^\circ-\Delta W\\
360^\circ& \rm{otherwise}.\\
\end{array}\right.
\end{equation}  
By testing several value of $\Delta W$, we finally set $\Delta W=90^\circ$. 
It is noted that such kind of model for $W_s$ is only an approximation, and it 
could lead to the discrepancy between the observation and simulation results.
So we need to use a better $W_s$ model in the future.  
Generally, the event source is near the 
solar equator, so the characteristic 
latitude of source location is set as $10^\circ$ north. Other important simulation 
parameters not varying are shown in Table \ref{tbl:param}.

\begin{threeparttable}[!htb]
\begin{centering}
\caption {Model Parameters Used in the Calculations.\label{tbl:param}}
\begin{tabular} {|l|l|l|} \hline
Parameter & Physical meaning & Value \\
\hline\hline
$E$ & Particles energy & $20$ MeV\\
\hline
$r_O$ & Observer solar distance & $1$ AU\\
\hline
$\Delta r$ & Shock space interval between two fresh injections& $0.001$ AU\\
\hline
$r_c$&Radial normalization parameter&$0.05$ AU\\
\hline
$\gamma$&Spectral index of source particles&$-3.5$\\
\hline
$\phi_c$ & Shock strength parameter & $15^{\circ}$ \\
\hline
$\lambda_\parallel$\tnote{a} &Particle mean free path&$0.16$ AU\\
\hline
$\kappa_\perp/\kappa_\parallel$\tnote{a} &Ratio between perpendicular and parallel diffusion
coefficient&$6.1$\%\\
\hline
$r_{b0}$ & Inner boundary & $0.05$ AU\\
\hline
$r_{b1}$ & Outer boundary & $50$ AU\\
\hline
\end{tabular}
\begin{tablenotes}\footnotesize
\item[a] For $20$  MeV protons in the ecliptic at $1$ AU.
\end{tablenotes}
\end{centering}
\end{threeparttable}

In order to investigate the relationship between solar wind speed $v_{SW}$ and 
CME speed $v_{CME}$, we obtain $v_{SW}$ observation data from $Wind$ spacecraft for 
the $204$ CME events to fit the relationship between $v_{CME}$
and $v_{SW}$. It is shown that $v_{CME}$ and 
$v_{SW}$ are positively correlated. As we assumed above that $v_s=v_{CME}$, the relationship between $v_{CME}$ and $v_{SW}$ would turn to that between $v_s$ and $v_{SW}$. Thus $v_{SW}$ can be represented by $v_s$ as 
\begin{equation}
v_{SW}=1.77\times10^{-5}{v_{s}}^2+425,\label{eq:vsw}
\end{equation}
here, $v_{SW}$ and $v_{s}$ are in the unit of km s${}^{-1}$. We also divide the events into 
several groups sorted on $v_{CME}$, and obtain the median values of $v_{CME}$ as the characteristic ones for each group.  So we obtain the counterpart values $v_{s}$ and $v_{SW}$ through the assumption above. And we use the 
characteristic ones in the simulations shown in Table \ref{tbl:simulations}.
The Table \ref{tbl:simulations} also shows the other input parameters in each simulation coming from the characteristic values
 of $v_{CME}$, $W_{CME}$ and source location longitude picked up from \citet{Kahler2013ApJ...769.110K} shown in Table \ref{tbl:kahler}.

\begin{table}[!htb]
\caption {Shock speed and width, and solar wind speed used in all simulations.
\label{tbl:simulations}}
\begin{tabular} {|c|c|c||c|c|c||c|c|c|} \hline
\multicolumn{9}{|c|}{Source location} \\ \hline
\multicolumn{3}{|c||}{N$10$W$48$} & \multicolumn{3}{c||}{N$10$W$77$} & \multicolumn{3}{c|}{N$10$W$115$}\\ \hline
$v_s$ (km/s)& $v_{SW}$ (km/s)& $W_s$ (${}^\circ$) & $v_s$ (km/s) & $v_{SW}$ (km/s) & $W_s$ (${}^\circ$) & $v_s$ (km/s) & $v_{SW}$  (km/s)& $W_s$ (${}^\circ$) \\ \hline
$450$ & $433.5$ & $167$ & $650$ & $433.5$ & $223$ & $620$ & $433.5$ & $190$\\\hline
$450$ & $433.5$ & $298$ & $650$ & $433.5$ & $261$ & $620$ & $433.5$ & $268$\\\hline
$450$ & $433.5$ & $360$ & $650$ & $433.5$ & $360$ & $620$ & $433.5$ & $360$\\\hline
$800$ & $455.4$ & $167$ & $1150$ & $455.4$ & $223$ & $900$ & $455.4$ & $190$\\\hline
$800$ & $455.4$ & $298$ & $1150$ & $455.4$ & $261$& $900$ & $455.4$ & $268$\\\hline
$800$ & $455.4$ & $360$ & $1150$ & $455.4$ & $360$& $900$ & $455.4$ & $360$\\\hline
$1175$ & $444.1$ & $167$ & $1450$ & $444.1$ & $223$ & $1325$ & $444.1$ & $190$ \\\hline
$1175$ & $444.1$ & $298$ & $1450$ & $444.1$ & $261$ & $1325$ & $444.1$ & $268$ \\\hline
$1175$ & $444.1$ & $360$ & $1450$ & $444.1$ & $360$ & $1325$ & $444.1$ & $360$\\\hline
$1600$ & $502.2$ & $167$ & $2100$ & $502.2$ & $223$ & $1750$ & $502.2$ & $190$ \\\hline
$1600$ & $502.2$ & $298$ & $2100$ & $502.2$ & $261$ & $1750$ & $502.2$ & $268$\\\hline
$1600$ & $502.2$ & $360$ & $2100$ & $502.2$ & $360$ & $1750$ & $502.2$ & $360$\\\hline
\end{tabular} 
\end{table}

\subsection{Simulation Output}

\begin{figure}[!htb]
\centering
   \includegraphics[width=11.0cm, angle=0]{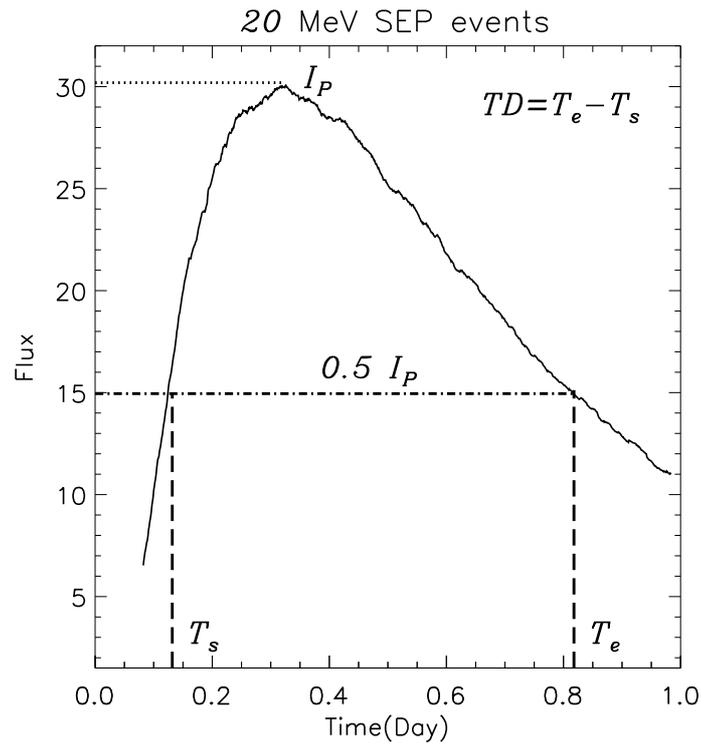}
\caption{Flux of $20$ MeV proton
during an SEP event with parameters shown in the text. The dotted line indicates
the peak intensity of the event, and the dash-dotted line indicates the half peak
intensity. $T_s$ and $T_e$ indicate the earliest and latest time when the intensity 
is half peak, respectively.
\label{Definition_fig1}} 
\end{figure}

For each data point, 3200000 virtual particles are calculated in our 
simulations. In our simulations, we obtain the time profiles of SEPs with characteristic speed 
and width of CME, with which we can get the SEP timescale, $TD$. For example, in
Figure \ref{Definition_fig1}, we show simulation results of $20$ MeV proton flux during 
an SEP event. In the simulation of Figure \ref{Definition_fig1}, we set solar wind speed 
as $502.2$ km s${}^{-1}$, longitude as $48$ degrees west, CME speed as $1600$ km s${}^{-1}$, 
CME width as $180^{\circ}$, other parameters are shown in Table \ref{tbl:param}.
In Figure \ref{Definition_fig1}, the dotted line indicates the peak intensity ($I_{p}$) 
of the event, and the dash-dotted line indicates the half peak intensity. $T_s$ and 
$T_e$ indicate the earliest and latest time when the intensity is half peak,
respectively. So we can obtain $TD=T_e-T_s$ from the time profile of intensity of simulation results.

From the results of the simulations we 
can also get the weighted averages as following. For example, in each range of shock
speed and longitude, we have three ranges of shock width, so we have three values of
$TD$ from simulation results with same shock speed and longitude but different 
shock width. For the three ranges of shock width we can get their percentage 
according to the number of events, with which the weighted value of $TD$ is obtained
from the individual values of $TD$.  

\begin{figure}[H]
\centering
   \includegraphics[width=11.0cm, angle=0]{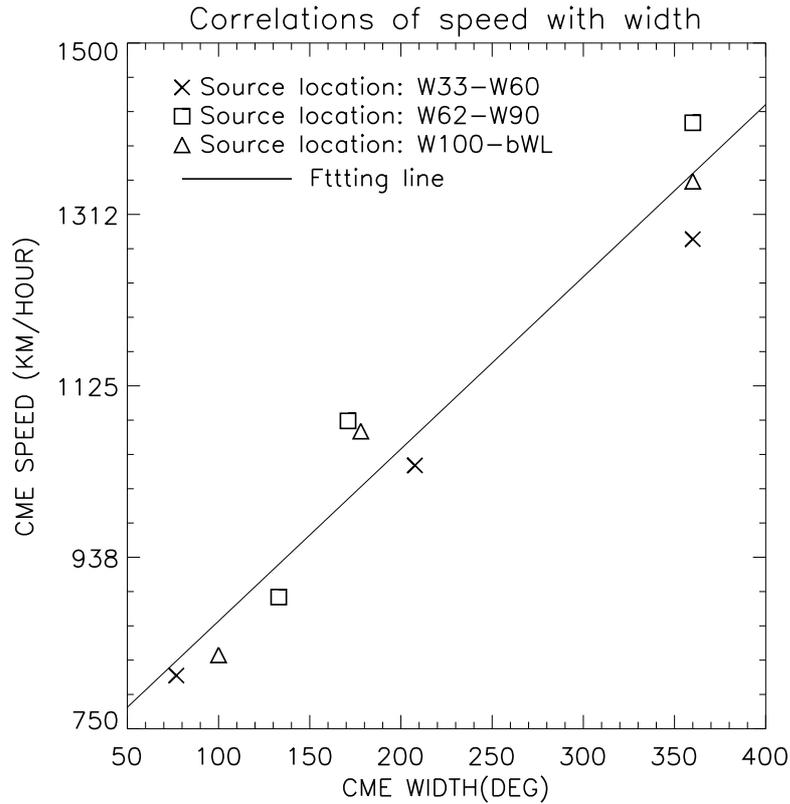}
\caption{Variations of CME speed as a function of
CME width with different source location. The crosses indecate CME speed averages of the CME width ranges in the source location range of W$33-$W$60$, and the squares are that of W$62-$W$90$, the triangles are that of W$100-$bWL. The line indicates fitting of the data. The symbols are
from observation data analysed by \cite{Kahler2013ApJ...769.110K}. 
\label{w_v_fit_fig2}}
\end{figure}

Further, we study the relationship between CME speed and CME width using the 
observation data in \citet{Kahler2013ApJ...769.110K}. We subdivide each longitude 
range into several groups sorted on CME width. We get average CME speed for
each group. The results are shown in Figure \ref{w_v_fit_fig2} as the relationship between 
average of CME speed and the median value of CME width. The three data points 
in each longitude group of Figure \ref{w_v_fit_fig2} match those of the three CME width bins 
of Table \ref{tbl:kahler}. The line indicates fitting of the data. It is found that in statistics the average CME width 
increases with the increasing of CME speed. 

\section{RESULTS}
\label{sect:results}
\begin{figure}[!htb]
\centering
   \includegraphics[width=14.0cm, angle=0]{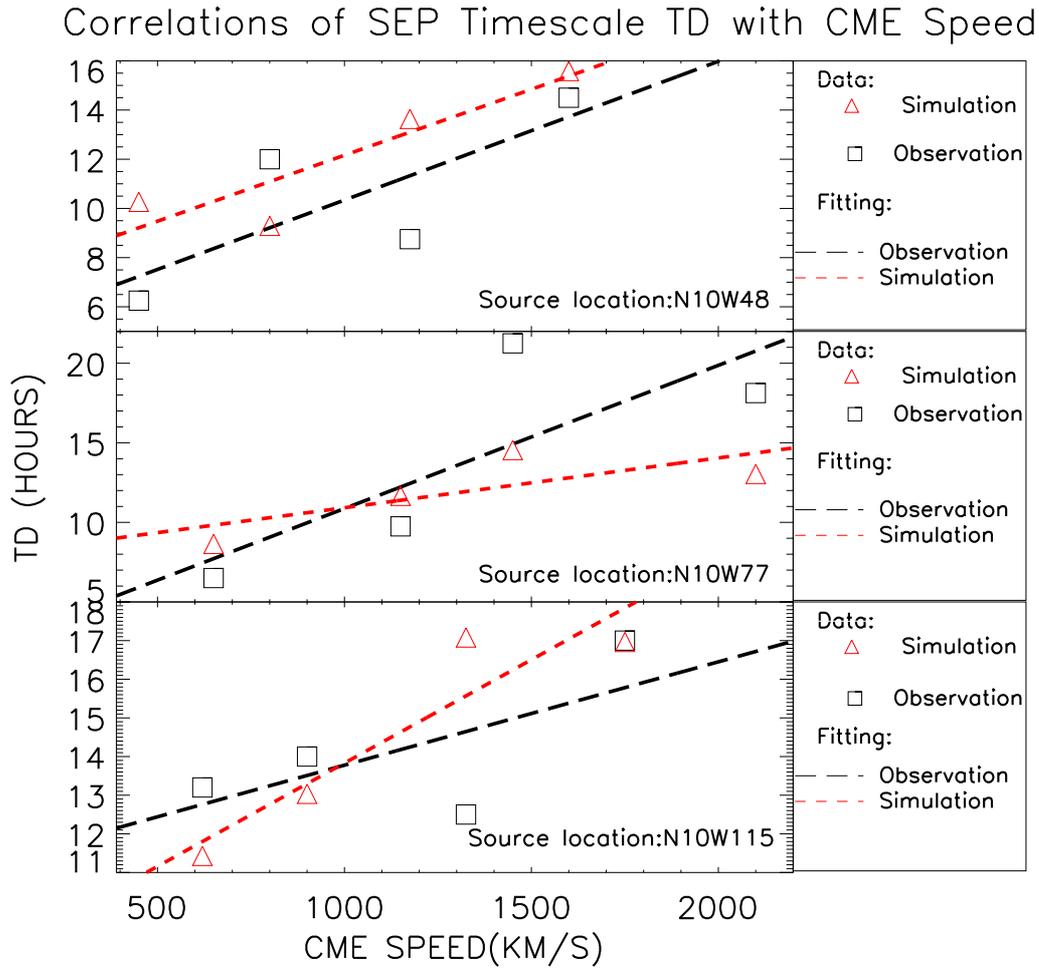}
\caption{SEP event timescale $TD$ vs. CME 
speed. Different panels indicate different source location. The black squares 
are from observation data analysed by \cite{Kahler2013ApJ...769.110K}. 
The $TD$ and CME speed for each data points correspond to those of 
Table \ref{tbl:kahler}. The red triangles indicate weighted 
average of simulation results. The black dashed lines indicate linear fitting 
of observation data. The red dashed lines indicate linear fitting of the weighted 
average of simulation results.
\label{td_v_obswei_fig3}}
\end{figure}

Figure \ref{td_v_obswei_fig3} shows SEP timescale $TD$ vs. CME speed for $20$ MeV SEP events detected at $1$ AU with different source locations in different pannels. The top, middle and bottom panels show different longitudes of source locations, $48^\circ$ west, $77^\circ$ west, and $115^\circ$ west, respectively. The black squares indicate spacecraft observation data in Table \ref{tbl:kahler} which are obtained from the data analysis of \cite{Kahler2013ApJ...769.110K}. The $TD$ and CME speed for each data points correspond to those of Table \ref{tbl:kahler}. The red triangles indicate the weighted average of simulations according to the distribution of number of events with different CME widths for any given CME speed interval obtained from the observation data in \citet{Kahler2013ApJ...769.110K} corresponding to the the abscissa of black squares. The red and black dashed lines indicate the linear fitting of the weighted average simulation results represented by the red triangles and that of the spacecraft observation data represented by the black squares, respectively. From Figure \ref{td_v_obswei_fig3} we can see, the simulation results show the similar trend of observation data, that is, the SEP timescale TD increases with CME speed.

\begin{figure}[!htb]
 \plotone{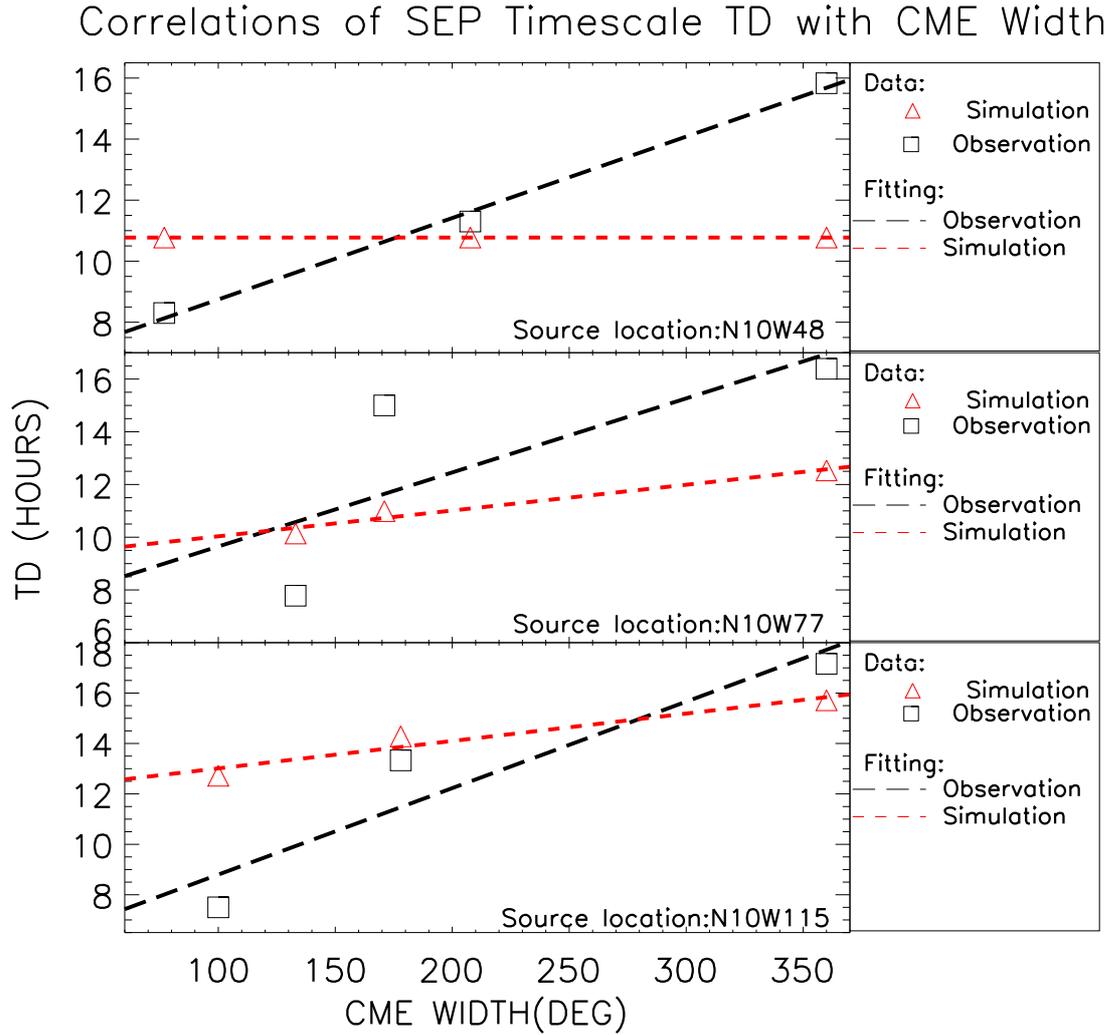} 
 \caption{SEP event timescale $TD$ vs. CME width. Different panels indicate different source location. The black squares are from observation data analysed by \cite{Kahler2013ApJ...769.110K}. The $TD$ and CME width for each data points correspond to those of Table \ref{tbl:kahler}. The red triangles indicate weighted average of simulation results. The black dashed lines indicate linear fitting of observation data. The red dashed lines indicate linear fitting of the weighted average of simulation results.
\label{td_w_obswei_fig4}}
\end{figure}

Figure \ref{td_w_obswei_fig4} shows plot similar as Figure \ref{td_v_obswei_fig3} except that x-coordinate is CME width. The value for each black squares correspond to those of Table \ref{tbl:kahler}. The red triangles indicate weighted average of simulation results according to the distribution of number of events with different CME speeds for any given CME width interval obtained from the observation data in \citet{Kahler2013ApJ...769.110K} corresponding to the abscissa of black squares. Similarly as in Figure \ref{td_v_obswei_fig3}, the 
red and black dashed lines indicate the linear fitting of the weighted average 
of simulation results and that of the spacecraft observation data, respectively. From Figure \ref{td_w_obswei_fig4} we can see, generally, the simulation results show the similar trend of observation data but with less slope, that is, the SEP timescale TD increases with CME width. However, from top panel of Figure  \ref{td_w_obswei_fig4} (N10W48) it is shown that, the observation results shows the SEP timescale TD increases with CME width, but the simulation results shows constant for different CME width. It is noted that our simulations could show deviation from observations due to modeling and statistical problems.

For further study on the contribution of CME speed and width to timescales separately, we plot the individual simulations and weighted average of simulation results as follows.

\begin{figure}[!htb]
\centering
   \includegraphics[width=14.0cm, angle=0]{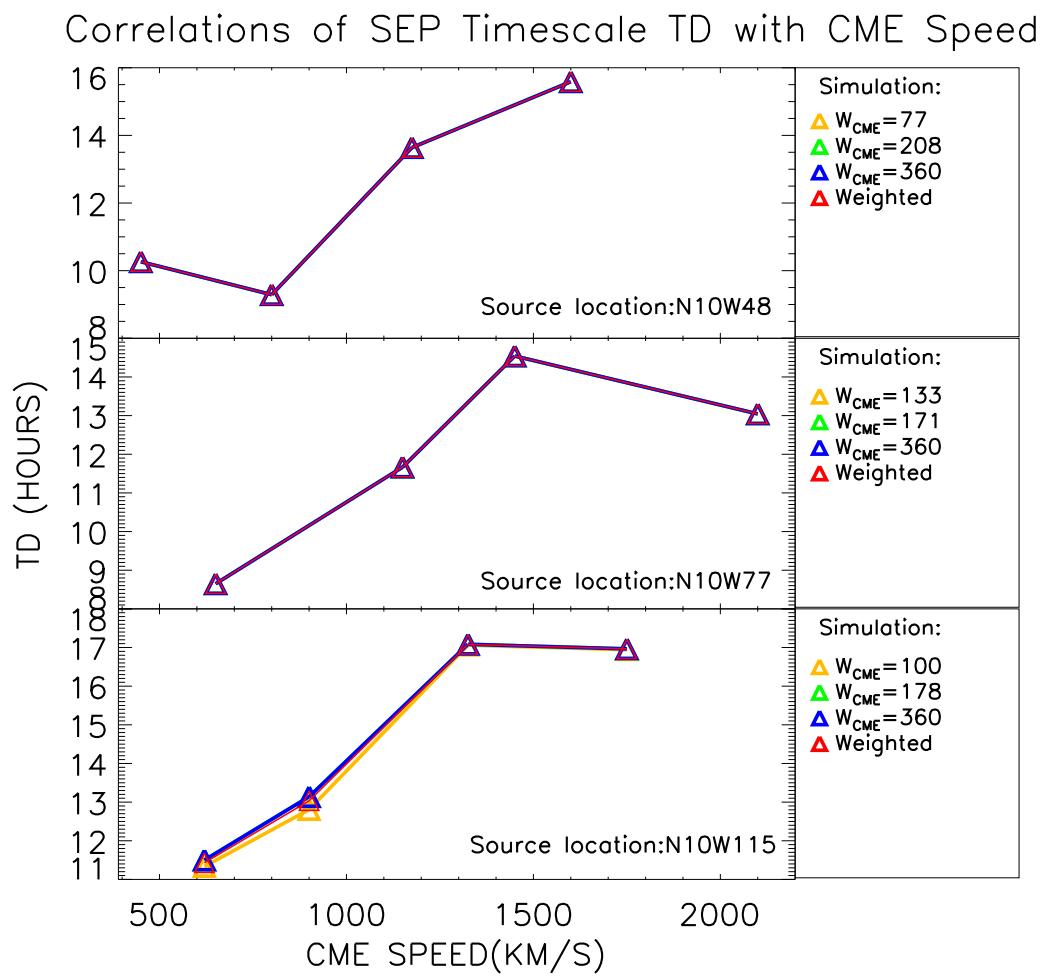} 
\caption{Simulations of SEP event timescale $TD$ vs. CME 
speed. Different panels indicate different source location. The yellow, green and blue triangles indicate simulations with different CME widths. The red triangles indicate weighted average of simulation results. 
\label{td_v_sim_fig5}}
\end{figure}

Figure \ref{td_v_sim_fig5} shows simulations of SEP timescale $TD$ vs. CME speed for $20$ MeV SEP events detected at $1$ AU with different source locations in different pannels. Similar as Figure \ref{td_v_obswei_fig3}, the top, middle and bottom panels show different longitudes of source locations, $48^\circ$ west, $77^\circ$ west, and $115^\circ$ west, respectively. The yellow, green, and blue triangles indicate simulations with different CME widths corresponding to those of Table \ref{tbl:kahler}. The each data point of the yellow line shows a individual simulation with a distinct CME speed but a common CME width $133^\circ$ and source location of N$10$W$48$, and so are the green and bule lines with other source location. Besides, the value of all data points are shown in Table \ref{tbl:simulations}. The red triangles indicate the weighted average of simulations according to the distribution of number of events with different CME widths for any given CME speed interval obtained from the observation data in \citet{Kahler2013ApJ...769.110K}.

From simulations of Figure \ref{td_v_sim_fig5} we can see, every single colored line increases, that is to say, for the same CME width, $TD$ generally increases with the increasing of CME speed, and the weighted average of $TD$ from simulations also generally increases with the increasing of CME speed. From the other aspect, the colored lines and symbols are almost overlap, from triangles with a common abscissa but different colors we can see, when CME speed is fixed, $TD$ with different CME widths are almost same, meanwhile, when CME width is fixed, $TD$ with different CME speeds are increased. So we suggest from our simulation that $TD$ is dependent on CME speed but not on CME width, which analysis of \cite{Kahler2013ApJ...769.110K} could not pick out.

\begin{figure}[!htb]
 \centering
   \includegraphics[width=14.0cm, angle=0]{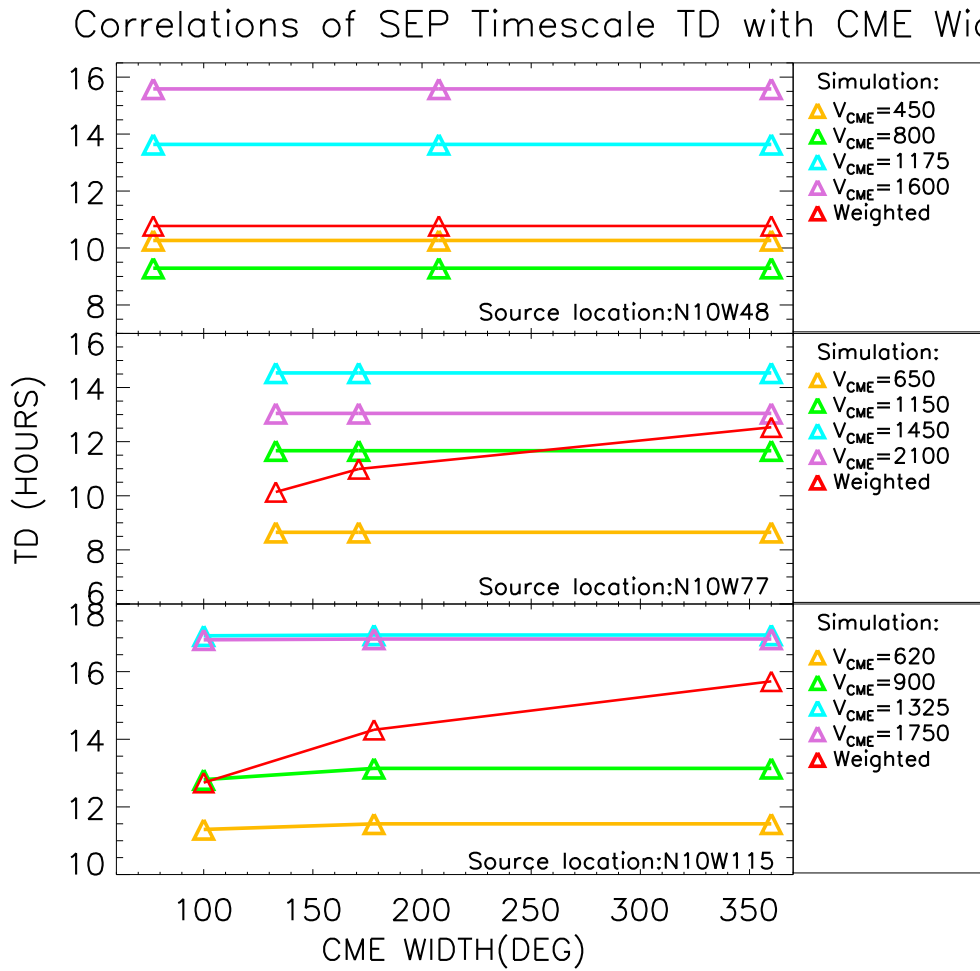}
 \caption{Simulations of SEP event timescale $TD$ vs. CME 
width. Different panels indicate different source location. The yellow, green, light blue, and purple triangles 
indicate simulations with different CME speed. The red triangles indicate 
weighted average of simulation results.
\label{td_w_sim_fig6}}
\end{figure}

Figure \ref{td_w_sim_fig6} shows plot similar as Figure \ref{td_v_sim_fig5} except that x-coordinate is CME width. The yellow, green, light blue and purple triangles indicate simulations with different CME speeds corresponding to those of Table \ref{tbl:kahler}. The each data point of the yellow line shows a individual simulation with a distinct CME width but a common CME speed $450$ km s${}^{-1}$ and source location of N$10$W$48$ , and so are the green and bule lines with other source location. Besides, the value of all data points are shown in Table \ref{tbl:simulations}. The red triangles indicate the weighted average of simulations according to the distribution of number of events with different CME speeds for any given CME width interval obtained from the observation data in \citet{Kahler2013ApJ...769.110K}.

From simulations of Figure \ref{td_w_sim_fig6} we can see, the yellow, green, light blue and purple lines are almost aclinic, that is to say, for the same CME speed, $TD$ generally keeps constant with different CME width. However, the red line which combines each individual line connecting data points of common CME speed simulations with weighted average increases, that is to say, for the same CME width, $TD$ increases with the increasing of CME speed. In addition, the weighted 
average of $TD$ increases with the increasing of CME width. The reason is that 
with larger CME width it is more likely that CME speed becomes larger, so the 
weighted average of $TD$ becomes larger consequently.

\section{CONCLUSIONS AND DISCUSSION}
\label{sect:discussion and conclusion}
Generally, the accurate measurement of the first arriving particles in SEP events
depends on the level of SEP flux background, so usually it is difficult to 
determine the timescales $TR$, $TO$, and $TO+TR$. However, $TD$, which indicates 
the duration of the SEP intensity above $0.5I_{p}$, has nothing to do with the 
first arriving particles, so the measurements of $TD$ are relatively accurate. 
Therefore, we only study the timescale $TD$, but do not study $TR$, $TO$, or 
$TO+TR$.

In this work, we use the SPTC to simulate the transport of SEPs assuming the ICME 
shock as a moving particle source with parameters obtained from spacecraft 
observations analysed by \cite{Kahler2013ApJ...769.110K},  and other parameters 
set as typical values of SEP events. From simulations we get SEP timescale $TD$ 
and compare with $TD$ values from spacecraft data analysis by 
\cite{Kahler2013ApJ...769.110K}. From spacecraft observations shown in 
\cite{Kahler2013ApJ...769.110K} we obtain the contribution of CME speed with the 
same CME width, and we also obtain that of CME width with the same CME speed. 
Finally, from simulation results of $TD$ we can obtain the average of $TD$ 
weighted with the observations contribution. 

Our simulations show that with the same CME speed, $TD$ keeps constant with the
increasing of CME width, but that the weighted average of $TD$ increases with the 
increasing of CME width. From spacecraft data analysis in 
\cite{Kahler2013ApJ...769.110K} it is shown that $TD$, which is actually weighted 
average, increases with the increasing of CME width. In addition, our simulations 
show that with the same CME width, $TD$ increases with the increasing of CME speed, 
and that the average of $TD$ increases with the increasing of CME width. It is also 
shown in \cite{Kahler2013ApJ...769.110K} with spacecraft data analysis that the 
weighted average of $TD$ increases with the increasing of CME speed. Our 
simulations generally agree with spacecraft observations data analysis of 
\cite{Kahler2013ApJ...769.110K} that the weighted average of $TD$ increase with both 
CME speed and width. Furthermore, with our modeling assumptions, our simulations 
show some results not shown in \cite{Kahler2013ApJ...769.110K} that $TD$ is 
dependent directly on CME speed, but independent on CME width.

In order to study whether TD increases with CME width or speed by using observation 
data, one should choose SEP events with same CME speed but different CME width to 
show if TD increases with CME width, and also one should choose SEP events with 
same CME width but different CME speed to show if TD increases with CME speed. 
\cite{Kahler2013ApJ...769.110K} did not do it because of limitation of events 
number. But simulations do not have this limitation, and that offers us physical 
insights behind the observations. We compare the weighted average of simulation to the result 
of \cite{Kahler2013ApJ...769.110K}, and we can show the trend of our weighted 
average generally agrees with the result of \cite{Kahler2013ApJ...769.110K}, so our 
work do not contradict the observation result of \cite{Kahler2013ApJ...769.110K}. 
Meanwhile, our individual results can be used to show if TD depends on CME width 
with same CME speed. 

The model we use to calculate flux includes many effects, such as the 
source, parallel and perpendicular diffusion, adiabatic cooling, etc., the overall 
effects could be very complicated, so we have to use numerical simulations to get 
the results. It is possible in some cases TD would decrease. But generally, TD has 
a trend to increase with the same CME width and increasing CME speed, and TD has a 
trend to be constant with the same CME speed and increasing CME width. Here, we 
compare the general trend between observations and simulations. 

We choose shock model conditions to favor larger particle injections with 
increasing speeds and widths in order to compare with observations. There are some 
parameters arbitrarily chosen and fixed in all simulations, we tried different 
parameters, for example, we tested simulations with different value of shock 
strength parameter $\phi_{c}$, such as $10^{\circ}$, $15^{\circ}$, $18^{\circ}$, 
and $25^{\circ}$, and we found they would not change our general results. In the 
future, we would continue to study the parameter effects in our model. 

The observational evidences of the first detected SEP onsets or releases associated with the
good magnetic connection to source were discussed in \cite{Ding2016RAA...16..8D}. Besides, \citet{RouillardEA11,RouillardEA12} suggested that SEP onsets could be
considered associated with the modeled first connections of field lines to shocks. 
On the other hand, \cite{QinAWang15} showed the onsets from SPTC simulation results 
can fit well with that from observations of $HELIOS$ $1$, $HELIOS$ $2$, and $IMP$ 
$8$ at different longitudes simultaneously with perpendicular diffusion. It is 
interesting to compare the effects of these models carefully in the future.

There are many authors working on numerical simulations to produce SEP profiles
from the shock onion shell model \citep[e.g.,][]{Verkhoglyadova2009APJ..693.894V, 
Verkhoglyadova2010JGR..115.A12103V, Wang2012ApJ...752..37W, QinEA13}, and they 
usually study the individual SEPs in detail, in this work, however, we are trying 
to study many SEPs with simulations so we can compare with observations 
statistically. CME width data from \cite{Kahler2013ApJ...769.110K} were observed 
by only one satellite, $SOHO$, so they are lack of determinacy. In the future, we 
would study the CME data of multi-spacecraft observations. In addition, we would 
study peak intensity of gradual SEP events associated with CMEs by comparing the 
simulations of SPTC with the spacecraft data analysis 
\citep[e.g.,][]{KahlerAVourlidas13}.

\begin{acknowledgements}
We are partly supported by grants NNSFC 41304135, NNSFC 41574172, NNSFC 41374177, 
and NNSFC 41125016, the CMA grant GYHY201106011, and the Specialized Research Fund 
for State Key Laboratories of China. The computations were performed by Numerical 
Forecast Modeling R\&D and VR System of State Key Laboratory of Space Weather 
and Special HPC work stand of Chinese Meridian Project. CME data were taken 
from the CDAW LASCO catalog, which is generated and maintained at the CDAW Data 
Center by NASA and The Catholic University of America in cooperation with the 
Naval Research Laboratory. $SOHO$ is a project of international cooperation
between ESA and NASA. We thank D. Reames for the use of the EPACT proton data.
\end{acknowledgements}


\label{lastpage}

\end{document}